\documentclass[sigconf,screen]{acmart}

\AtBeginDocument{%
  \providecommand\BibTeX{{%
    \normalfont B\kern-0.5em{\scshape i\kern-0.25em b}\kern-0.8em\TeX}}}

\copyrightyear{2025}
\acmYear{2025}
\setcopyright{rightsretained}
\acmConference[DIS '25]{ACM Designing Interactive Systems Conference (DIS) 2025}{July 5--9, 2025}{Funchal, Madeira}
\acmBooktitle{Proceedings of the ACM Designing Interactive Systems Conference (DIS) 2025, July 5--9, 2025, Funchal, Madeira}
\acmISBN{XXXX}
\usepackage{xcolor}

\begin{document}

\title[Designing for Community Care]{Designing for Community Care: \\
Reimagining 
Support for Equity \& Well-being in Academia}


\author{Beatriz Severes}
\orcid{0000-0001-8751-1783}
\author{Ana O. Henriques}
\orcid{0000-0002-8326-1630}
\affiliation{%
 \institution{ITI/LARSyS, University of Lisbon}
 \city{Lisbon}
 \country{Portugal}}

 
\author{Rory Clark}
\orcid{0009-0000-6775-5035}
\affiliation{%
 \institution{Bristol University}
 \city{Bristol}
 \country{United Kingdom}}

\author{Paulo Bala}
\orcid{0000-0003-1743-0261}
\affiliation{%
 \institution{ITI/LARSyS}
 \city{Funchal}
 \country{Portugal}}

\author{Anna R. L. Carter}
\orcid{000-0002-2436-666X}
\affiliation{%
\institution{Northumbria University}
\city{Newcastle}
\country{UK}}

\author{Rua Mae Williams}
\orcid{0000-0002-6182-8923}
\affiliation{%
\institution{Purdue University}
\city{West Lafayette}
\country{US}}

\author{Geraldine Fitzpatrick}
\orcid{0000-0002-2013-2188}
\affiliation{%
\institution{\href{http://www.geraldinefitzpatrick.com}{geraldinefitzpatrick.com}}
\city{Vienna}
\country{Austria}}

\renewcommand{\shortauthors}{Severes et al.}

\begin{abstract}
Academic well-being is deeply influenced by peer-support networks, yet they remain informal, inequitable, and unsustainable, often relying on personal connections and social capital rather than structured, inclusive systems. Additionally, institutional well-being responses frequently focus on student populations, neglecting the emotional labour of faculty and staff, reinforcing an exclusionary academic culture. Drawing on HCI methodologies, participatory design, and care ethics, this workshop will provide a space for rethinking how academic communities can support inclusive networks. Through pre-workshop engagement, co-design activities, and reflection, participants will examine systemic gaps in networks and explore ways to embed care, equity, and sustainability into academic peer-support frameworks – from informal, exclusionary models to structured, inclusive care-based ecosystems.
At the end of the workshop, participants will co-develop design strategies for integrating care and resilience in academic ecosystems, resources for designing equitable support systems, and a peer network invested and committed to fostering a supportive academic community.

\end{abstract}




\maketitle

\section{Introduction and Background}

Community support is fundamental for academic success and well-being, fostering resilience, collaboration, and career development while addressing personal needs and mitigating isolation. These networks help reduce burnout and empower individuals to navigate challenges, creating an inclusive environment where academics can thrive both personally and professionally~\cite{Pyhalto2017, Ronkkonen2023, Hollywood2020}.
However, academic support networks often remain informal and inequitable despite their benefits, leaving many researchers excluded and under-served~\cite{Miner2019}. This is driven by systemic pressures and unaddressed inequities that contribute to widespread mental health challenges, particularly among Early Career Researchers (ECRs)~\cite{Kozlowski2022, Nicholls2022}.

ECRs, including graduate students, postdoctoral researchers, and newly appointed investigators, are six times more likely to experience anxiety and depression than the general population~\cite{Evans2018, Moss2022} due to unique vulnerabilities from academic pressures, financial insecurity, and blurred boundaries between professional and personal life~\cite{Hollywood2020, Jackman2022}. Systemic infrastructural inequalities --- such as unequal access to funding, resources, and opportunities~\cite{Wilkins-Yel2019, Mcgee2019, Ymous2020}, inadequate support for disabilities~\cite{Ymous2020}, pervasive harassment~\cite{OBrien2016}, or racial discrimination~\cite{Mcgee2019} --- create an environment that exacerbates stress and undermines the well-being and professional potential of researchers~\cite{OBrien2016, Wilkins-Yel2019}. This environment is particularly challenging for marginalised groups, such as women and disabled or racialised academics~\cite{Harrington2023, Wilkins-Yel2019, Camacho2011, Harlow2003}, who face heightened burdens due to epistemic violence that undermines their role as credible contributors~\cite{Ymous2020}. This occurs in numerous forms of discrimination~\cite{OBrien2016} such as microaggressions~\cite{Wilkins-Yel2019, Mcgee2019}, group exclusion~\cite{Wilkins-Yel2019, Miner2019}, or delegitimisation of expertise due to disability, gender and race~\cite{Wilkins-Yel2019, Ymous2020}.

This marginalisation extends beyond a lack of resources. It constitutes a systemic devaluation of marginalised scholars’ lived experiences and expertise~\cite{Bjrn2022, Washington2020}. This devaluation not only limits material access but also actively dismisses their knowledge~\cite{Williams2023} --- effectively excluding these scholars from shaping technologies and academic discourse, which in turn compounds daily challenges and restricts opportunities for meaningful engagement. This further exacerbates stress and undermines well-being and professional potential~\cite{Casad2021, Charles2022, Kozlowski2022, Wilkins-Yel2019, Williams2023}. 
As such, establishing formal, equitable support structures that incorporate feminist care ethics \textit{ethos} is essential to redistribute power and promote interdependence~\cite{Tronto1990, Tronto2013, Held2005, Henriques2024, Henriques2025}, creating a more inclusive academic environment.

As academics progress into faculty roles, these challenges often intensify, requiring them to balance mentorship, publishing, teaching and administrative duties while providing emotional support to students and junior colleagues, frequently acting as first responders to mental health crises without training and shouldering significant emotional labour that exacerbates burnout~\cite{Price2017, Meeks2023, Butler2022}. Unfortunately, institutional responses to mental health often focus disproportionately on student well-being, neglecting the needs of faculty and staff~\cite{Butler2022}. Stigma, lack of awareness, and inadequate accommodations further limit access to resources, perpetuating a culture that prioritises individual achievement over collective care~\cite{Browning2017, Nicholls2022, Carter24}. The absence of consistent, structured, and intentional support systems underscores the urgent need for inclusive and resilient frameworks to address these systemic issues.

Indeed, academic support is most effective when community networks move beyond deficit models to foster interdependence~\cite{Williams2023}. Grounding support structures in care ethics, isolated shortcomings are reframed as collective practices that benefit the whole community~\cite{Toombs2015, Toombs2017}. By recognising and building on strategies that use technology and social networks, they move beyond simply “fixing” perceived deficits~\cite{Williams2023, Ymous2020}. Through this lens, effective support is inherently interdependent~\cite{Toombs2017}, as reciprocal care is a shared responsibility that enhances everyone’s well-being~\cite{Held2005}, reinforcing the need for structured support systems that acknowledge these dynamics~\cite{Toombs2017, Carter24}. Although academic narratives often champion self-reliance, community maintenance research reveals that social connections and care are essential for sustainable progression and well-being~\cite{Toombs2017, Ymous2020}. By adopting feminist care–based support frameworks, institutions can counter deficit-focused approaches~\cite{Ymous2020} and enable academic communities to build collective resilience and shared responsibility, creating environments where relational care is the foundation of well-being~\cite{Toombs2017, Williams2023, Toombs2015}.

While informal peer-support systems can foster meaningful connections and a sense of belonging, they are inherently inequitable and unsustainable, relying heavily on pre-existing relationships, personal networks, and social capital, which often exclude marginalised groups and fail to address systemic challenges~\cite{Kozlowski2022, Casad2021}.
To address these inequities, academia must shift from relying on informal mechanisms to adopting formalised peer-support structures rooted in care ethics, providing a refined lens for examining the relational dynamics behind effective support networks~\cite{Toombs2015}. This approach moves beyond a transactional view of care to reveal how everyday acts of support --- from sharing expertise to providing emotional support --- contribute to community cohesion. By emphasising the reciprocal nature of care~\cite{Toombs2017}, this approach demonstrates that meaningful support emerges from interdependent practices rather than isolated efforts~\cite{Toombs2015}, underscoring the crucial role of relational labour in sustaining both academic and social communities. By formalising care-based peer-support structures, academic systems can shift the focus from isolated deficits to shared strengths and mutual support, institutionalising reciprocal practices and prioritising collective well-being over individual efforts. As such, this approach emphasises relational accountability, shared responsibility, and adaptability, reframing academic well-being as a collective commitment rather than an individual burden~\cite{Brown2024, Murnane2018}. 

The integration of care ethics into Human-Computer Interaction~(HCI) practices emphasises the need to address the emotional labour, power dynamics, and systemic inequities inherent in academic life. HCI research has demonstrated the value of amplifying community voices to co-create equitable and effective peer-support systems~\cite{Im2024, Agapie2022, Tachtler2021}. Drawing on insights from HCI, feminist care ethics, digital civics, and participatory design practices, this approach encourages the academic community to reflect on these challenges and explore possibilities grounded in lived experiences. 

Building on these principles, this workshop aims to address the gap between informal support networks and structured, institutional frameworks, creating more accessible systems that embed care, resilience, and collaboration into academic ecosystems. By centering the lived experiences of the academic community, we seek to foster a culture of belonging and collective care~\cite{Nicholls2022, Woolston2020}. 

As such, this workshop will:
\begin{enumerate}
    \item Provide an inclusive space for participants to collaboratively explore how principles of equity, care, and value-sensitive design can inform the creation of resilient and sustainable academic peer-support networks, and
    \item Facilitate the co-creation of actionable guidelines for care-based academic community-building through hands-on activities and interdisciplinary collaboration.
\end{enumerate}

Drawing on participants’ lived experiences, the workshop’s outputs will result in an resource kit designed to create inclusive, sustainable academic communities that empower participants to build equitable support systems and promote collective care, contributing to a reflection on reimagining academic well-being and transforming academic ecosystems. 
\section{Motivation}

Academic communities often rely on informal yet meaningful peer-support systems, as researchers face a number of systemic challenges that compromise their well-being and hinder professional development. However, the ad hoc nature of these support structures often leaves many without consistent support. In response, this workshop will discuss what is needed to shift from these informal mechanisms to intentionally designed, inclusive care-focused networks that prioritise equity, community care, and sustainability. 
By moving from informal, chance-based systems to structured, equitable frameworks grounded in care ethics, intersectionality, and sustainability, we aim to reimagine academia as a resilient, collaborative ecosystem where academic well-being is a collective responsibility.
Participants will co-create artifacts and guidelines for academic peer support informed by their lived experiences, addressing systemic barriers and exploring alternative ways to embed equity and care into academic culture.
The workshop aims to generate practical tools, spark conversations that challenge existing norms, and inspire a shift toward more inclusive and supportive practices within academia. By bringing together diverse voices from across academic roles and disciplines, this workshop seeks to catalyse broader cultural change, inviting participants to collaboratively design solutions that go beyond individual resilience and laying a foundation for academia to expand as a more equitable, interconnected, and sustainable community.

\section{Goals \& Questions}

Our workshop leverages care ethics principles to reimagine academic support networks as sustainable and inclusive ecosystems. Through collaborative activities, participants will identify key community-building values and develop actionable strategies to integrate these principles into academic environments.

Our primary goal is to \textbf{co-develop guidelines for community support that prioritise equity and care through collective care and collaboration while addressing systemic barriers faced by academic communities}. To achieve this, activities will focus on:

\begin{enumerate}
    \item identifying shared values essential for academic community-building
    \item translating these values into actionable strategies, and 
    \item fostering cross-disciplinary and cross-role collaboration.  
\end{enumerate}

The workshop centers on three fundamental questions: 1) Which values best support inclusive, resilient academic communities?; 2) How can these values be operationalised into strategies for long-term sustainability?; and 3) How can participants critically examine their roles, privileges, and responsibilities in designing sustainable academic support networks? Discussions surrounding these questions will engage participants to co-develop inclusive and actionable guidelines for care-based academic support structures.

\section{Workshop Structure}
\textbf{\textit{Format:}} One-day, in-person workshop.
This workshop is designed as a collaborative space where participants will critically examine and reimagine academic peer support. Through structured activities grounded in feminist care ethics, collective well-being, and resilience, participants will be encouraged to co-create academic structures that move beyond informal, ad hoc support toward sustainable, care-centered support systems. By reflecting on their lived experiences, articulating core values, and prototyping actionable strategies, we will collectively explore how academic support can shift from individual burden to shared care-based responsibility~\cite{Tronto1990}.

\subsection{Pre-Workshop}
\textbf{\textit{Ice-breaker:}} Prior to the workshop, participants will receive a welcome email detailing the agenda, objectives, accessibility needs, and essential event information. They will be invited to join a dedicated online platform (e.g., Discord) where they can engage in reflective prompts and informal discussions to begin exploring their experiences with academic peer support. Participants will be encouraged to share short posts or examples of their networks and support systems, fostering early connections and setting the stage for in-depth discussions during the workshop.

These pre-workshop activities aim to build a sense of shared purpose while ensuring accommodations are met and avoiding cognitive overload.

\subsection{During the Workshop}
The workshop will span a full day, structured into a morning session focused on foundational discussions and an afternoon dedicated to hands-on activities and the development of community-driven interventions.

\textbf{\textit{Welcome} (10min):} Organisers will introduce the goals, structure, and guiding principles of the workshop, emphasising our feminist care ethics \textit{ethos} and core values of equity, resilience, and allyship toward community-building~\cite{Henriques2024, Henriques2025, Wilkins-Yel2019b, Wilkins-Yel2019, Williams2023, Ymous2020, Carter24}.

\textbf{\textit{Keynote and Q\&A} (30mins):} Introductory keynote with Dr. Fitzpatrick to ground the day’s discussions and set the tone for subsequent activities.

\textbf{\textit{Participant Introductions} (30mins):} Participants will each have up to three minutes to introduce themselves, discuss their experiences with academic community-building, and share aspirations for more sustainable peer-support structures.

\textbf{\textit{From Lived Experience to Core Values} (45mins):}
Participants will work in small groups (3-5 people) to map their experiences with academic peer support, identifying key moments, challenges, and informal structures they have relied upon. Groups will then collaboratively extract the core values that have shaped these experiences, considering what principles are essential for inclusive, sustainable, and resilient academic communities. This discussion anchors our exploration of participants’ lived realities while establishing a shared foundation for subsequent activities.

\textbf{\textit{Break} (15mins)}

\textbf{\textit{From Values to Action} (1h30mins):} Building on the core values identified, participants will collaboratively transform these principles into tangible strategies and interventions for academic peer support. This structured process moves from reflection to action, grounding care ethics in real-world applications.

\textit{Part 1. Defining} (30mins): Participants select a core value and explore how it manifests (or fails to manifest) in academic contexts. This discussion will be guided by questions such as ‘how does this value show up in academic spaces today?’, ‘what systemic barriers hinder its realisation?’, and ‘what would this value look like in a well-supported peer community?’

\textit{Part 2. Prototyping} (1h): Groups will create speculative prototypes that operationalise these values --- examples might include alternative support structures, policies, tools, or community practices. As such, participants will be encouraged to develop concrete, care-centered interventions using design-centered techniques like design fictions or zine-making to explore this implementation. 

\textbf{Lunch (1h30mins)}: Participants are encouraged to continue informal discussions with the organisers and peers.

\textbf{\textit{From Actions to Sustainable Structures} (1h30min):} Participants will refine their prototypes, shifting from conceptual interventions to building structured guidelines and actionable community-building strategies.

\textit{Part 3. Synthesis} (45mins): Groups will map out the real-world applications of their proposed interventions, detailing what specific actions can be taken to make this idea work, what tools, resources, or policies would support its implementation, or what barriers might arise, and how can they be addressed.

\textit{Part 4. Establishing Guiding Principles} (45mins): Participants will then distill their work into adaptable community-building guidelines, focusing on inclusivity, sustainability, and resilience, as well as the other core values established by each group. These outputs will form the basis of post-workshop collaborations and potential institutional recommendations (see Section \ref{Expected-Outcomes}).

\textbf{Break (30mins)}

\textbf{\textit{Group Reflection} (45mins):} Each group will present their strategies and identify three essential features for care-based academic support structures. Participants will collectively discuss, rank, and refine these elements, ensuring shared ownership of outcomes.

\textbf{\textit{Wrap-up} (15mins):} Closing thoughts and next steps.
\subsection{Post-Workshop} Workshop outputs will be compiled and shared via the same online platform where we gathered for Pre-Workshop activities. There, participants can continue discussions, refine ideas, and explore potential applications and outcomes. Additionally, we intend for this platform to establish the beginning of a community-support structure that we will nurture and hopefully grow as a care-focused space for academic peer-support.

\section{Expected Outcomes} \label{Expected-Outcomes} With participant consent, insights will be synthesised and systematised toward resource-sharing and knowledge building. We plan on submitting a report to the SIGCHI Equity Committee to foster future implementation of the workshop's outputs to ensure a more long-term and wider impact of these discussions. Additionally, the workshop will lay the groundwork for a collaborative position paper to be published as part of a broader edited collection or special issue (e.g., an Interactions article or DIS provocation) in order to support participants, as well as our broader academic communities in applying workshop insights to their own academic environments. We also plan to compile all participant submissions into arXiv proceedings and conduct future workshops or engagements --- such as World Cafe activities\footnote{\url{https://theworldcafe.com/key-concepts-resources/world-cafe-method}} or SIGs --- to further extend these discussions.

\section{Intended Audience }
We will gather 10-20 participants whose work or experiences resonate with the themes and objectives of our workshop.

This workshop is open to individuals across academia and related fields committed to fostering inclusive and equitable peer-support systems, in particular:

 → \textbf{Academic Stakeholders,} including early career researchers, senior researchers, mentors, and faculty members committed to creating community-driven support systems; 

 →  \textbf{Equity and Inclusion Advocates,} dedicated to addressing systemic inequities in academic institutions, policymakers, and scholars in HCI, Feminist Ethics, and Digital Civics who promote participatory, collaborative practices. 


We actively encourage diverse participation, especially from individuals representing historically underrepresented or marginalized communities, whose unique insights – often overlooked – are essential for expanding our collective understanding and enhancing the impact of our workshop outcomes.

\subsection{Accessibility}

Pre-workshop activities will be offered in both synchronous and asynchronous formats, allowing flexibility for participation across time zones and accommodating different needs. 
Detailed instructions for each activity will be available on the Miro board, enabling participants to join and engage at their own pace before, during, and after the workshop.
We are dedicated to creating a supportive and inclusive environment where everyone can fully participate. To achieve this, we will share a pre-workshop form for participants to request specific accommodations, which we will provide to the best of our ability.

\section{Call for Participation}
Building on our ongoing work in academic \textbf{well-being, care ethics, community-building}~\cite{Severes2024, Henriques2024, Henriques2025, Carter24, Ashcroft2025, Tachtler2021, Fitzpatrick2016, Ymous2020, Williams2023}, we will leverage our professional, institutional, and personal networks to disseminate and attract a diverse group of participants. 
Additionally, a dedicated workshop website will serve as a central hub for information. This will include the call for participation, submission guidelines, and workshop goals, as well as instructions for applying, which will be shared in May to accommodate the early bird deadline.

We encourage submissions in various formats --- including design prototypes, exploratory research, work-in-progress, position statements, auto-ethnographies, and case studies --- that explore different aspects of designing equitable and sustainable academic peer-support networks. Main topics of interest include the role of design in fostering inclusive and accessible community support networks; interventions that promote connectivity, peer-support, and resilience within communities; ethical considerations, challenges and strategies for design and implementation of support systems; case studies of community support networks; theoretical frameworks that inform the design and evaluation of community support systems.


Participants are invited to submit either a 1-page reflective statement (PDF) or a 2-minute video (MP4) outlining their experiences with academic community-building or their interest in fostering inclusive peer-support systems. We welcome alternative formats to accommodate diverse communication styles. With the authors’ consent, accepted contributions will be shared on our website, as arXiv proceedings, and among participants. Any submission may be withdrawn immediately upon request.
\section{Why DIS’25?}
Given the intrinsic relationship between academic systems, design, and the collective well-being they foster, it is essential to reflect on how we approach the structures that shape research communities. Design, through its capacity to reimagine relationships and shared environments, becomes a means to envision alternative futures rooted in care and equity. Through this workshop, we seek to embody DIS 2025’s vision of leveraging ecological metaphors, such as resilience, adaptability, and interdependence, to inspire more sustainable and supportive academic cultures. 
By framing academic peer-support networks as interconnected ecosystems --- resilient, adaptable, and inclusive --- this workshop builds on design-driven tools for collaboration and mutual care to address systemic inequities and promote the intentional design of equitable academic environments.

This workshop aligns with the spirit of DIS by adopting participatory approaches to rethinking support systems and embedding care ethics and sustainability into the heart of academic culture. Highlighting co-creation as a core principle, we emphasise designing not for individuals alone but for the collective, echoing the conference’s broader and central themes of interconnectedness. Through this lens, we seek to expand resilient community support by integrating social, technological, and institutional dimensions, while leveraging design methods to redefine academic support for its members and promote inclusive, equitable, and enduring communities.

\bibliographystyle{ACM-Reference-Format}
\bibliography{bibliography/references_clean}

\end{document}